# Recent Investigations of Cascaded GEM and MHSP detectors

R. Chechik, A. Breskin, G. P. Guedes, D. Mörmann, J. M. Maia, V. Dangendorf, D. Vartsky, J. M. F. Dos Santos, and J. F. C. A. Veloso

*Abstract*
on detectors comprising cascaded gas electron multipliers (GEM) and cascaded GEMs with micro-hole and strip plate (MHSP) multiplier as a final amplification stage. We discuss the factors governing the operation of these fast radiation-imaging detectors, which have single-charge sensitivity. The issue of ion-backflow and ion-induced secondary effects is discussed in some detail, presenting ways for its suppression. Applications are presented in the fields of photon imaging in the UV-to-visible spectral range as well as x-ray and neutron imaging.

## I. Introduction

THE gas electron multiplier (GEM) [1] presents attractive features, making it an amplifying-element of choice in a variety of radiation imaging detectors [2]. It is fast, robust and can be produced with relatively large area (30cm× 30cm). The GEM is collecting and multiplying charges induced by radiation either in a gas volume or on a solid radiation converter. The multiplied charges are collected onto a patterned anode, to provide high-resolution localization. According to the size of the initially-deposited charge, the GEM can be employed as a stand-alone multiplier or in a cascaded mode. The optical opacity of the electrode and the confinement of the avalanche to the holes result in highly suppressed photon-mediated and photon-induced processes. Consequently, high charge gain, of typically $10^3$-$10^4$ is achieved with single GEM in common counting gases [3]. Cascaded triple- and quadruple-GEMs attain gains of about $10^5$-$10^6$ in various Ar-based noble gas mixtures [4] and higher gains in Ar-$CH_4$ and in pure $CF_4$ [5]. Multi-GEM detectors guarantee stable operation in charged-particles [6] and x-ray [7] detection, and high sensitivity to single charges, the latter is the basis for multi-GEM gaseous photomultipliers (GPMT) [4], [8]. They are investigated and applied for particle tracking [9], as imaging elements in TPCs [10], for x-ray [11]-[12] and thermal-neutron imaging [13]-[14] and for the imaging of UV photons [15] and visible light [16], [8].

In this article we explain some processes governing the operation of radiation detectors based on multi-GEMs and multi-GEMs followed by the novel micro-hole and strip (MHSP) multiplier [17]-[18]. We present our recent results on ion backflow and its reduction or suppression in both cascaded multipliers. Applications to x-ray, neutron and photon imaging are presented.

## II. Electron Transport in Cascaded-GEM Detectors – A Concise Recall

The coupling of cascaded-GEMs to gaseous converters is straight forward, and the efficiency of electron focusing into the first-GEM holes has been thoroughly studied [19]. It was established that full detection efficiency for charges deposited above the GEM is obtained with small field $E_{drift}$ in the conversion region, of the order of 100-500V/cm at atmospheric pressure.

The coupling of GEM to a solid converter may be done in two modes. The first is *semitransparent mode*, in which the converter is placed at some distance above the first GEM, with a field $E_{drift}$ between them. The second is the *reflective mode* (fig. 1), in which the converter is deposited on top of the first GEM, and a field $E_{drift}$ is defined in the gap above it, by a mesh electrode. In many cases the reflective mode is preferable, permitting the use of a thick converter with high conversion efficiency.

Manuscript received October 29, 2003. This work was supported in part by the Israel Science Foundation, by the Planning and Budgeting Committee of the Council for Higher Education in Israel, and by the Fundação para a Ciência e a Tecnologia (FCT), Lisbon, Portugal, through project POCTI/FNU/49553/02 of the Instrumentation Centre (unit 217/94), Physics Department, University of Coimbra.

R. Chechik is with the Weizmann Institute of Science, 76100, Rehovot, Israel (corresponding author, telephone: +972-8-9344966, e-mail: Rachel.chechik@weizmann.ac.il).
A. Breskin is with the Weizmann Institute of Science, 76100, Rehovot, Israel.
G. P. Guedes was with the Weizmann Institute of Science, 76100, Rehovot, Israel, on leave from the Lab. for Nucl. Instrum. COPPE/UFRJ, Rio de Janeiro, Brazil. He is now with the Universidade Estadual de Feira de Santana, Av. Universitária, BR 116-Norte, Km 03, 44030-460, Feira de Santana-BA, Brazil
D. Mörmann is with the Weizmann Institute of Science, 76100, Rehovot, Israel.
J. M. Maia was with the Weizmann Institute of Science, on leave from University of Coimbra and University of Beira-Interior, Portugal. He is now with the University of Coimbra and University of Beira-Interior, Portugal.
V. Dangendorf was with the Weizmann Institute of Science, 76100, Rehovot, Israel, on leave from PTB, Braunschweig, Germany. He is now with the PTB, D-38116 Braunschweig, Germany.
D. Vartzky is with Soreq NRC, Yavne 81800, Israel.
J. M. F. Dos Santos is with the University of Coimbra, Portugal.
J. F. C. A. Veloso is with the University of Coimbra and University of Aveiro, Portugal.

It is important to note that electrons emitted from a solid converter into gas experience energy-dependent scattering processes, and possible loss due to backscattering into the converter ((1) in fig. 1). The electron loss depends on the gas type and on the electric field at the converter surface; it is most significant in noble gases and at small fields [20]. A field >1 kV/cm in atmospheric pressure is required to minimize the backscattering loss, which may cause inefficient electron focusing into the GEM holes. We extensively investigated the issue for semitransparent [21]-[22] and reflective [23]-[24] operation modes. In the *semitransparent mode*, we confirmed that the highest possible ratio of the field $E_{GEM}(1)$ (inside the GEM1 holes) to $E_{drift}$ should be kept. The efficiency may not be 100% unless GEM1 gain is high, in the order of $10^3$. In the *reflective mode* we showed that a sufficiently high field on the converter surface is established by high GEM1 voltage [23]-[24]. We confirmed that fully efficient focusing ((2) in fig. 1) and detection of the electrons is obtained when $E_{drift}=0$ and $E_{GEM}(1)$ is high, corresponding to GEM1 gain of $10^2$ to $10^3$, depending on the gas [23]-[24].

The transport of electrons between successive GEM electrodes may also be understood from fig. 1: the extraction of electrons from the $n^{th}$ GEM towards its subsequent is expected to increase with the ratio $E_{trans}(n)/E_{GEM}(n)$, but their focusing into the holes of the next GEM decreases with the ratio $E_{trans}(n)/E_{GEM}(n+1)$. The process has been comprehensively studied [19], confirming the expected trends. Preferable operation conditions are therefore based on a compromise, with a large fraction of the electron charge (~50%) lost on the GEMs faces ((3) and (4) in fig. 1).

It is important to bear in mind that maintaining high transparency for the electrons implies similarly high transparency for avalanche ions flowing along the same field-lines in the opposite direction, with very little diffusion.

### III. ION BACKFLOW

Back-flowing avalanche ions may cause physical and chemical damage to the solid converter, and may affect its performance by charge accumulation on the surface. Furthermore, with converters of efficient secondary electron emission, secondary avalanches are initiated (*ion feedback*), severely limiting the detector gain. Ions back flowing into the drift volume ((5) in fig. 1) affect the operation of TPC devices as well [25]. The quantity named here *ion backflow* is the fraction of total avalanche-generated ions reaching the converter. Previous studies [19], [26]-[28] of this important quantity provided the general understanding of its dependence on the GEM geometry, gas type, pressure, fields, etc. It was demonstrated that the most effective parameter for ion backflow reduction is the field $E_{drift}$. The ion-charge flow could also be reduced by blocking it in the one-before-last GEM (e.g. using single-conical GEM, GEM with small holes, smaller GEM voltage) [29], and compensating for the gain loss by higher gain on the last GEM.). But this asymmetric operation mode is limited by GEM instability at high voltages.

With *gas converters* $E_{drift}$ can be maintained small, and thus limiting the ion backflow to the few % level. However, with *solid converters*, in order to minimize electron backscattering, the field at the converter surface should always be kept high, and therefore the ions are strongly attracted to it. We have recently studied the ion backflow in a 4-GEM gaseous photomultiplier (with 4 identical GEM electrodes) with a *reflective CsI photocathode* [30]. As is obvious from fig. 1, the ion extraction and transport follow similar trends as the electrons: ion transmission from the $n^{th}$ GEM upwards increases with the ratio $E_{trans}(n-1)/E_{GEM}(n)$, but their focusing into the holes of GEM(n-1) decreases with the ratio $E_{trans}(n-1)/E_{GEM}(n-1)$. By varying all possible fields in the 4-GEM detector, but without sacrificing full photoelectron extraction and detection efficiency (namely keeping high surface field on GEM1 and total gain of $10^6$), we showed that the ion backflow may be reduced at best to ~20% (fig. 2). Operating the device with asymmetric powering of the last two GEMs may slightly further reduce the ion backflow, by a few %. We demonstrated an additional reduction of the ion backflow, to 10% (fig. 2), by operating the induction gap (between the last GEM and the anode electrode) under high field $E_{ind}$, in parallel-plate multiplication mode, with gain $>10^2$. In this case the final avalanche takes place at the induction volume where the majority of ions are produced and the high ratio $E_{ind}/E_{GEM}(4)$ guarantees their inefficient focusing back into the last GEM. But in this approach the readout anode (fig. 1) is no more decoupled from the multiplication electrodes, and there is a distinct ion component in the signal captured on the anode.

The MHSP electrode (fig. 3) developed recently [17]-[18], opens a way for further reduction of ion backflow. This GEM-like electrode has anode- and cathode- strips etched on its bottom face, and it operates in double-stage multiplication: the avalanche produced inside the hole is further multiplied on the anode strips. Consequently, a significant part of the ions are collected on adjacent cathode strips and, possibly, on the mesh or patterned readout electrode placed below the MHSP (fig. 3). We have recently shown [31] that by replacing GEM4 in the detector of fig. 1 with a MHSP electrode, new operating conditions could be used: the last mesh acts as a cathode, the charge flow is partially blocked by small $E_{trans}(3)$ and the gain-loss is compensated by the anode-strips multiplication. Under such conditions we reduced the ion backflow to ~2%.

To practically suppress the ion backflow we have introduced [30] an active pulsed-gate electrode, as suggested in [25]. It consists of a parallel-wires plane, uniformly biased in the *open state* and *alternately biased*, by a +/- voltage pulse, in the *closed state*. We have confirmed full electron transmission in the *open state*. We incorporated the gate electrode between GEM3 and GEM4 in a 4-GEM detector with *reflective CsI photocathode*, and operated it at 100 torr $CH_4$ in order to deliberately enhance the ion-induced secondary effects from the photocathode. Using 80V, 10μsec-long gate pulses, triggered by the anode electron signal, and counting the secondary ion-induced pulses (fig. 4), we showed [30] that the gate reduces the ion backflow by a factor $10^4$, as shown in fig. 5. We confirmed identical gate operation at atmospheric pressure, with gate pulses of 150-200 V. The dramatic ion backflow suppression is of coarse achieved at the expense of counting rate capability, due to the ~10μsec dead time, which could be reduced by optimizing the detector's geometry.

We clearly demonstrated the need for such efficient suppression of ion backflow, in a multi-GEM photon detector with semitransparent K-Cs-Sb photocathode of visible-light range sensitivity. For this photocathode we measured the secondary electron emission probability under impact of back-flowing ions in $Ar/CH_4$ mixtures, to be 0.05 to 0.5 electrons/ion, the higher probability was recorded in pure Ar. Without ion gating, we found that secondary avalanches induced by impact of ions from the first GEM, limit its gain (fig. 6 and 7); the gain limit varies from 30 to 1000, depending on the $CH_4$ percentage in the mixture. This dependence is derived mainly from the differences in electron backscattering and in ion backflow. The present interpretation, based on our recent systematic measurements, replaces our previous one given in [8].

## IV. Timing and Localization

The use of solid radiation converters coupled to gaseous electron multipliers results in improved timing and localization properties [32]. Surface-emitted electrons undergo almost-synchronous fast multiplication within the GEM holes; the small gaps between consecutive GEMs prevent diffusion-governed avalanche spread, resulting in very fast pulses [4]. For example, in a 4-GEM detector operated in $CF_4$ with a reflective CsI photocathode, we recorded time resolutions of ό=0.33 and 1.6 nsec, with 150 and single photoelectrons, respectively. The point-like conversion and the small electron diffusion result also in excellent localization resolution while the granularity of the first GEM-electrode is expected to contribute typically less than 30 ìm rms to the localization precision; the final avalanche lateral size is of the order of the hole diameter.

Multi-GEM detectors currently demonstrate localization resolutions of a few tens of ìm rms with charged particles and soft x-rays; these are obtained with costly highly-integrated readout electronics [7]. A different challenge in radiation imaging with large-area GEM-based detectors is to find more economic readout methods. For that purpose the initial, spatially narrow, charge induced on the readout electrode should be broadened, either by introducing a wide induction gap (between the last GEM and the readout anode) [33], or by recording induced charges behind a resistive charge-collection anode [34]. The latter is efficiently broadening the charge distribution according to the distance between the resistive layer and the readout electrode, which is capacitively coupled and conveniently grounded. We investigated both techniques and applied them to the imaging of UV light, soft x-rays and fast neutrons, as discussed below.

## V. Applications of Cascaded-GEM Detectors

### A. Detectors of visible light

Photon detectors with semitransparent K-Cs-Sb photocathode and 4 standard Kapton GEMs are currently sealed in packages filled with atmospheric $Ar/CH_4$ (95:5). The deposition & sealing system, the processes developed at our laboratory as well as detector details are described in [8]. Sealed detectors of that type had so far a lifetime of the order of one month, with quantum efficiency above 6% at 360nm [8]. The latest version of this device includes a gate electrode, to suppress ion-feedback. Despite the apparent compatibility of Kapton GEMs, at least on a short term, efforts are taken to produce GEM electrodes of ultrahigh vacuum compatible materials such as ceramic and silicon [35]. Other candidate electron multipliers could be cascaded MCP-like glass-capillary multipliers [36].

### B. UV-photon detectors

Among the important advantages of a reflective photocathode deposited on the first GEM (fig. 1) is its operation with $E_{drift}=0$, which renders the detector insensitive to ionizing background radiation crossing the drift volume. We have proposed [24] to use such reflective-CsI photon detectors for the imaging of UV photons in Ring Imaging Cherenkov (RICH) devices, where high-multiplicity ionizing radiation is a principal limiting factor. We demonstrated the operation in pure $CF_4$ [5], which permit conceiving windowless detectors with the radiator and the photon-detector operating within the same gas volume. Detectors based on this principle are currently investigated for a Hadron-blind Cherenkov detector for the PHENIX experiment at RHIC [37].

Based on the possibility to operate GEM- and GEM/MHSP-cascades in noble gas mixtures [4], we foresee their application in gas scintillation proportional chambers, with the scintillation volume directly coupled to the UV detector, within the same gas volume. Stable operation, with high gain and good energy resolution, was presented recently with a single MHSP element in atmospheric Ar/Xe(5%) mixture [38]. Xe-operated multi-GEM UV-detectors with CsI photocathodes were recently proposed by the XENON collaboration for recording scintillation light from liquid-Xe WIMP detectors [39]. Recent results on the operation of cascaded GEMs in such two-phase (liquid-gas) cryogenic detectors are given in [40].

*C. X-ray imaging detectors*

We have developed a 3-GEM x-ray imaging detector, where x-rays are converted in a gas gap. It has a 400-μm pitch striped X-Y anode-readout plane, coupled to fast (2.1 nsec/tap) discrete-element delay lines [12]. With a broad induction-gap we matched the lateral charge-distribution width, in $Ar/CO_2$ (70:30) and in $Ar/CH_4$ (95:5), to the anode pitch. The induced-pulse time-width was matched to the frequency transmission-characteristics of the line, with <20% signal attenuation along 100mm [33], [12]. Despite the x-ray conversion in gas, we demonstrated intrinsic localization resolutions of σ ~70 μm. This simple and fast delay-line based readout scheme permits photon-counting imaging at 100kHz.

The resistive electrode approach was recently investigated with cascaded GEM and GEM/MHSP multipliers, coupled to a Wedge-&-Strip (W&S) [41] readout electrode, of 1.6 mm pitch. Localization resolution of σ = 60 μm was recorded with 6 keV x-rays and a 3-GEM multiplier; a resolution of σ = 100 μm was recorded with single UV photons and a 4-GEM multiplier coupled to a CsI photocathode. In both cases the W&S electrode was places behind a resistive anode. In a 2-GEM/MHSP detector the ion-induced signals were recorded at the W&S electrode placed behind a resistive cathode; a resolution of σ =100 μm was obtained with 6 keV x-rays. More details are given in reference [38].

We currently foresee the development of high-resolution cascaded multi-GEM secondary electron emission (SEE) imaging detectors [32] with CsI-coated x-ray converters.

*D. Thermal neutron imaging*

We are currently developing a fast novel thermal-neutron imaging-detector [42], schematically depicted in fig. 8. It comprises a thin (<0.5 mm), $^{10}$B-rich (22% boron content by weight) liquid-scintillator, with emission peaked at 350 nm, matched to the quantum efficiency of the photon detector described above (V.*A*). The detector is designed to have 0.95 efficiency to thermal neutrons (of 1.8 Å) and only $3\times10^{-3}$ efficiency to γ. With pulse-shape discrimination procedure these efficiencies become 0.8 and $2\times10^{-5}$, respectively. The position resolution obtained with a fiber faceplate window is ~500 μm and the time resolution (determined by the liquid converter thickness) is ~200 ns.

*E. Fast-neutron imaging*

We have recently developed a novel fast-neutron imaging detector, depicted schematically in fig. 9. The detector prototype (10cm × 10cm) comprises a 1mm thick polyethylene converter coupled, via 1 mm thick gas converter, to a 3-GEM detector. Protons originating from neutron conversion in the foil are converted in the gas and imaged with the multi-GEM detector. We implemented the resistive layer approach discussed above and delay-line readout [43]. The readout-electrode has non-overlapping X-Y pads of 2mm pitch, at the backside of a resistive layer anode; the pads are coupled to a fast delay-line (2.7 ns/tap). We showed that a resistive layer with surface resistivity >30MΩ/□ (e.g. ~160 nm Ge evaporated on glass or on 3.2 mm G-10) has ~94% signal transmission. We recently demonstrated fast-neutron imaging capability with localization resolution < 1 mm rms. Fig. 10 shows a radiographic image of carbon cylindrical phantoms and a metal wrench (the wrench's screw is clearly visible at the center of the field view), recorded with a broad-energy neutron beam (3 to 10 MeV) and small statistics. The efficiency of the detector to MeV neutrons is ~0.2%. To achieve higher efficiencies, for practical application in fast neutron radiography, it is foreseen to combine 25 identical layers in sequence, to obtain ~5 % efficiency. In a recent experiment with a pulsed fast-neutron beam a time resolution of σ = 1.7 ns was obtained.

## VI. SUMMARY

We have extensively studied the properties of multi-GEM detectors, particularly the processes involved in electron and ion transport and the key parameters controlling them. We have shown that with solid converters (semitransparent or reflective) ion backflow cannot be safely reduced below 15 to 20% of the total charge without affecting the electron transport, namely losing quantum efficiency or gain. With the new MHSP multiplier replacing the last GEM in the cascade, ion backflow could be reduced to 2%. We demonstrated that a pulsed gating-electrode introduced in the cascade could effectively suppress ion backflow (and the resulting ion-induced feedback) by a factor of $10^4$, without losing detection efficiency, although with introduction of some dead time.

We are applying multi-GEM detectors for the fast and efficient imaging of soft x-rays, fast- and thermal- neutrons and UV-to-visible photons. Some examples were shortly presented, with concise discussion of readout methods as well as localization and timing properties.

## VII. ACKNOWLEDGMENT

V. Dangendorf and D. Mõrmann are grateful to the support of MINERVA foundation; G. Guedes thanks the Brazilian grant agency CAPES for financial support; J. M. Maia and J. F. C. A. Veloso acknowledge support from FCT, Lisbon, Portugal. A. Breskin is a W. P. Reuther Professor of Research in the Peaceful use of Atomic Energy.

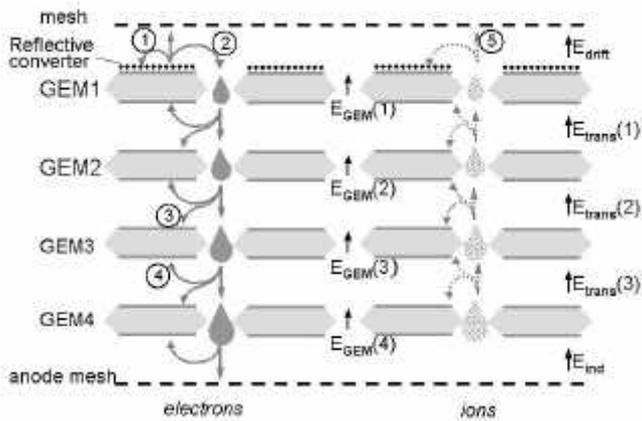

Fig. 1. Definition of notations and scheme of electrons (left) and ions (right) transport processes in a 4-GEM detector with reflective solid converter.

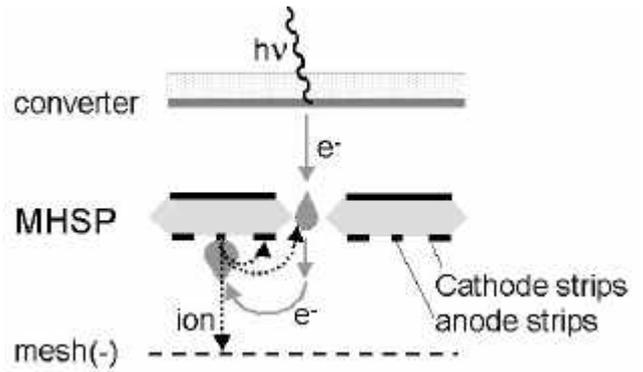

Fig. 3. The micro-hole and strip plate multiplier is a GEM-like device, etched with anode and cathode strips at its bottom. Two-stage electron multiplication occurs in the holes and further at the anode strips. Most of the final avalanche-ions may be collected at the nearby cathode strips and on the cathode mesh below.

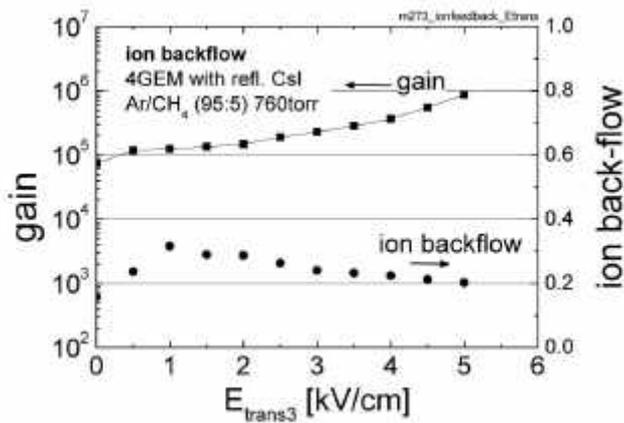

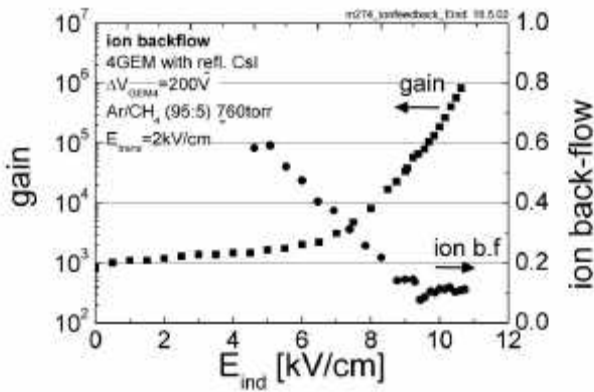

Fig. 2. Optimization of ion backflow in a 4-identical-GEMs detector, with reflective converter; on top is shown the effect of the transfer field between GEM3 and GEM4 and at the bottom the effect of the induction field chosen in multiplication regime. The best ion backflow is 20% and 10%, respectively, of the total charge gain.

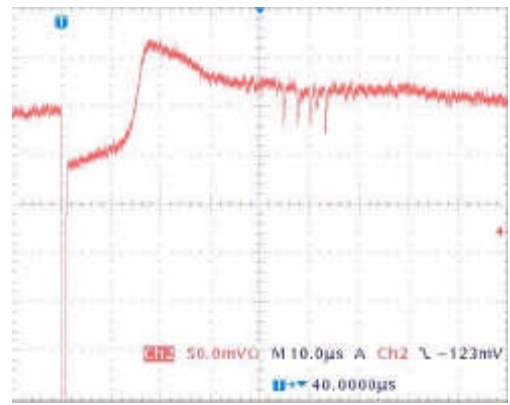

Fig. 4. Fast pulses recorded from the gated 4-GEM detector, with gate open. Starting at 45 ìsec after the main avalanche, corresponding to the ion drift-time to the converter, ion-induced secondary avalanches are seen.

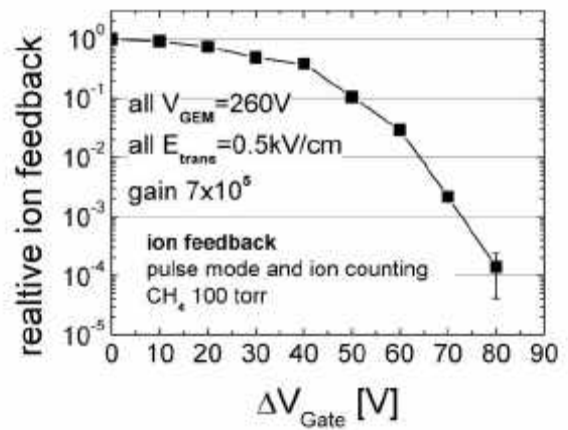

Fig. 5. Relative ion feedback obtained from ion-induced feedback avalanches counting (fig. 4), as function of gate-pulse voltage. Suppression by factor $10^4$ is demonstrated here at low pressure with 80V gate-pulse; identical performance was confirmed at atmospheric pressure with 150-200V gate pulse.

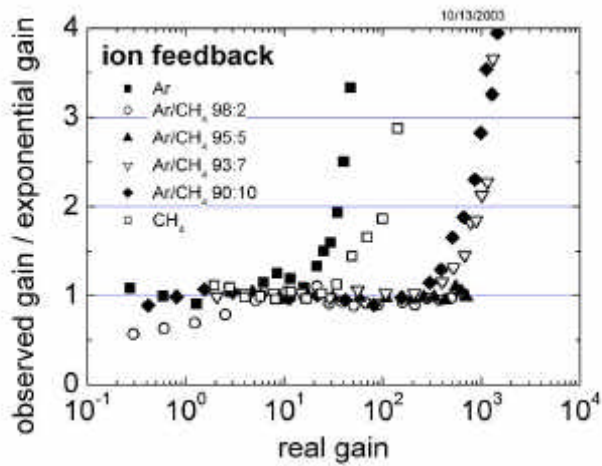

fig. 6. Deviation of the recorded gain from the exponentially-fitted one, as function of the last, measured in current mode with a single-GEM and semitransparent K-Cs-Sb photocathode, in various gas mixtures. The deviation, which occurs at different gains in different gases, is due to intensive secondary avalanches, shown in fig. 7.

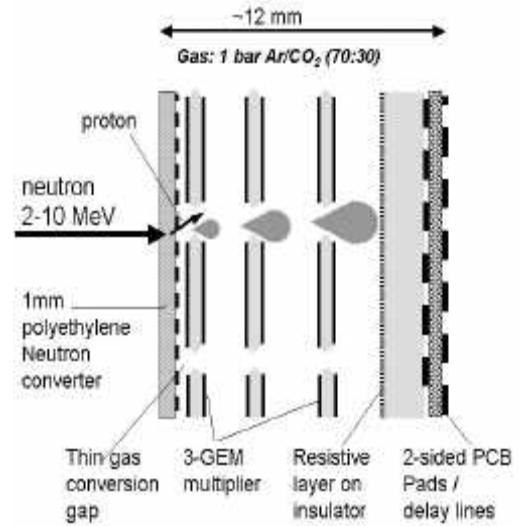

Fig. 9. Schematic view of the fast-neutron detector: 2 to 10 MeV neutrons are converted in polyethylene and the resulting protons are detected and imaged with a 3-GEM detector; spatially-broadened charge signals are recorded behind the resistive layer, on X-Y anode pads connected to delay-line readout.

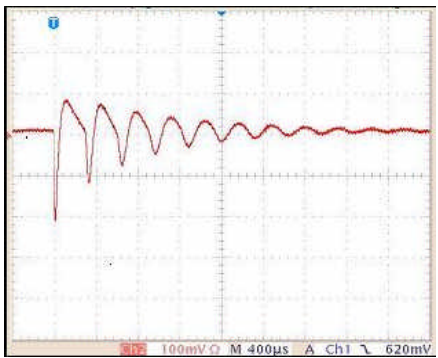

Fig. 7. Ion-induced feedback pulses, observed with a semitransparent K-Cs-Sb photocathode and a single GEM, operated with pure argon at gain >40.

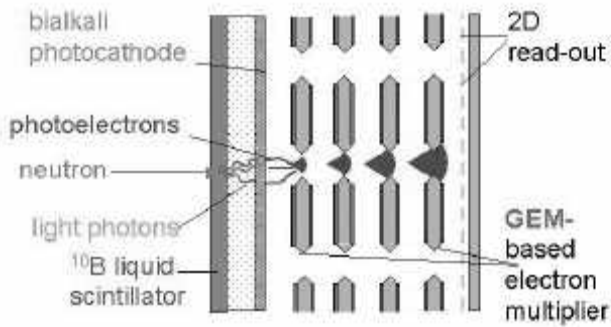

Fig. 8. A scheme of the novel thermal-neutron detector currently under development; it is based on boron-rich liquid scintillating converter coupled to a sealed multi-GEM photon detector. The thin scintillator, the efficient light coupling and the fast photoelectron multiplication in the GEM-based multiplier, guarantee fast response, low sensitivity to gamma background and very high neutron detection efficiency.

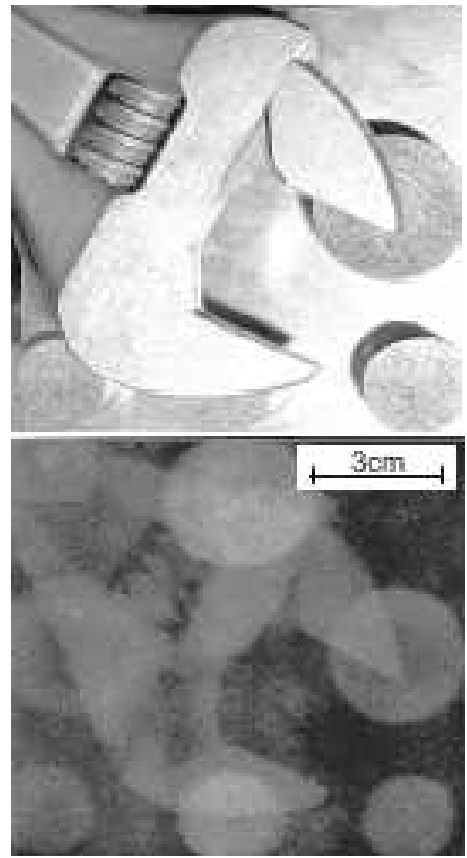

Fig. 10. A radiographic image (bottom) of 6 cylindrical carbon phantoms and a metal wrench (top), recorded with a broad– energy neutron beam (3-10 MeV) and the novel fast-neutron detector.